\documentclass{iau_FM}
\usepackage{graphicx}

\newcommand\rhodm{\rho_{\rm dm}}
\newcommand\rperi{r_{\rm peri}}
\newcommand\rhohf{\rho_{150}}
\newcommand{\Msun}{{\rm M}_\odot}
\newcommand{\kpc}{\,{\rm kpc}}
\newcommand{\Gyr}{\,{\rm Gyr}}
\newcommand{\pc}{\,{\rm pc}}
\newcommand{\cm}{\,{\rm cm}}
\newcommand{\GeV}{\,{\rm GeV}}
\newcommand\dd{{\rm d}}

\title[Dark matter in the Fornax, Sculptor and Leo I dwarf spheroidal galaxies] 
{Dark-matter central density, annihilation $J$-factor and decay $D$-factor of the Fornax, Sculptor and Leo I dwarf spheroidal galaxies}

\author[C.\ Nipoti, R.\ Pascale \& J.~M.\ Arroyo-Polonio]   %% give here short author list %%
{Carlo Nipoti$^1$,
Raffaele Pascale$^2$
  \and José María Arroyo-Polonio$^{3,4}$}

\affiliation{$^1$Dipartimento di Fisica e Astronomia ``Augusto Righi'', Universit\`{a} di Bologna, \\via Gobetti 93/2, 40129, Bologna, Italy, email: {\tt carlo.nipoti@unibo.it} \\[\affilskip]
  $^2$ INAF – Osservatorio di Astrofisica e Scienza dello Spazio di Bologna, \\via Gobetti 93/3, 40129 Bologna, Italy\\[\affilskip]
  $^3$Instituto de Astrofísica de Canarias,\\ Calle Vía Láctea s/n E-38206 La Laguna, Santa Cruz de Tenerife, España\\[\affilskip]
  $^4$Universidad de La Laguna,\\ Avda. Astrofísico Francisco Sánchez E-38205 La Laguna, Santa Cruz de Tenerife, España
}

\pubyear{2024}
\jname{~}
\editors{~}

\begin{document}

\maketitle

%{\em \tiny November 4, 2024}

\begin{abstract} 

The dwarf spheroidal galaxies (dSphs) satellites of the Milky Way (MW)
are nearby astrophysical laboratories to study the nature of dark
matter (DM). We present some properties of the DM halos of the three
classical dSphs Fornax, Sculptor and Leo I, obtained using dynamical
models based on distribution functions depending on the action
integrals. In particular, we report accurate estimates of their
central DM density $\rhohf$ (measured at a distance of $150\pc$ from
the galaxy centre), which is relevant for galaxy formation studies and
for models of self-interacting DM, and their DM annihilation
$J$-factor and decay $D$-factor, which are key tools for indirect
DM detection experiments. Among these three galaxies, Fornax has the
highest $J$- and $D$- factors (but the lowest $\rhohf$), while Leo I
has the highest $\rhohf$ (but the lowest $J$- and $D$- factors).

\begin{flushright}
{\em November 4, 2024}
\end{flushright} 
\end{abstract}

\firstsection % if your document starts with a section,
\section{Introduction}

The dwarf spheroidal galaxies (dSphs) are gas-poor
low-surface-brightness faint stellar systems that can be studied in
detail only in the Local Group.  The kinematics of the Local Group
dSphs reveals that these galaxies are dominated by dark matter (DM)
down to their centre (\cite[Battaglia \& Nipoti 2022]{Bat22}). These
nearby dwarf galaxies are thus ideal astrophysical laboratories to
study the properties of DM.

The characterization of the DM halos of dSphs is important not only
for our understanding of galaxy formation and evolution, but also to
explore the nature of DM.  Here we focus on three quantities that are
widely used in dSph-based studies of DM. The central DM density
$\rhohf$, defined as the density at a distance of $150\pc$ from the
galaxy centre, is often considered as a benchmark to constrain galaxy
formation models in different frameworks and especially in
self-interacting DM cosmologies (e.g.\ \cite[Correa 2021]{Cor21}).
The so-called $J$-factor and $D$-factor are the astrophysical
normalization terms that are necessary to interpret measurements of
(or upper limits on) $\gamma$-ray flux from dSphs in terms of particle
physics properties of the DM annihilation and decay, respectively, and
are thus key tools for indirect DM detection experiments.

For sufficiently distant, spherically symmetric targets, the $J$- and
$D$-factors can be computed, respectively, as (\cite[Evans,  Sanders \& Geringer-Sameth 2016]{Eva16})
\begin{equation}
  J(\theta) = \frac{2\pi}{d^2}\int_{-\infty}^{+\infty}\dd z\int_0^{\theta d}\rhodm^2(R,z)R\dd R,
\label{eq:jfactor}  
\end{equation}
and 
\begin{equation}
  D(\theta) = \frac{2\pi}{d^2}\int_{-\infty}^{+\infty}\dd z\int_0^{\theta d}\rhodm(R,z) R\dd R,
\label{eq:dfactor}
\end{equation}
where ($R$,$z$) are cylindrical coordinates with origin in the galaxy
centre and $z$ along the line of sight, $d$ is the distance of the
galaxy, and $\theta=R/d$.  Here we focus on measurements with $\theta =
0.5^{\circ}$, corresponding approximately to the angular resolution of
the Fermi-LAT telescope in the GeV range.

In this paper, combining results from different studies that we have
carried out with our collaborators over the last few years, we report
measurements of $\rhohf$, $J$-factor and $D$-factor of three MW
classical dSphs: the Fornax, Sculptor and Leo I dSphs.  Throughout
this work we assume the following distances: $d=138\kpc$ for Fornax
(\cite[Saviane, Held \& Bertelli 2000]{Sav00}; \cite[Pascale et
  al.\ 2018]{Pas18}), $d=83.9\kpc$ for Sculptor
(\cite[Mart{\'\i}nez-V{\'a}zquez et al.\ 2015]{Mar15};
\cite[Arroyo-Polonio et al.\ in prep.]{Arr_inprep}) and $d=256.7\kpc$
for Leo I (\cite[Held et al.\ 2010]{Hel10}; \cite[Pascale et
  al.\ 2024]{Pas24}).

\section{Dynamical models of the Fornax, Sculptor and Leo I dSphs}

\begin{table}
  \begin{center}
    \caption{Estimates of the DM parameters $\rhohf$, $J$-factor and
      $D$-factor of the Fornax, Sculptor and Leo I dSphs. The $J$- and
      $D$-factors are measured within $0.5^\circ$ and reported in the
      standard units with masses expressed in $\GeV$. For each
      quantity we report the median and the $1\sigma$ interval (16th
      and 84th percentiles).}
  \label{tab:darkpar}
 {\scriptsize
  \begin{tabular}{|l|c|c|c|}\hline 
    {Galaxy} & {\bf $\rhohf$ $\left[10^8\Msun\kpc^{-3}\right]$} & {\bf $\log\,J(0.5^\circ)$ $\left[\GeV^2\cm^{-5}\right]$} & {\bf $\log\,D(0.5^\circ)$ $\left[\GeV\cm^{-2}\right]$} \\
    \hline 
    Fornax   & $0.339_{-0.029}^{+0.024}$ & $18.36_{-0.09}^{+0.06}$ & $18.55_{-0.05}^{+0.03}$ \\
 \hline
    Sculptor   & $1.02_{-0.11}^{+0.15}$ & $18.14_{-0.11}^{+0.11}$ & $18.08_{-0.09}^{+0.09}$ \\ \hline    
    Leo I   & $3.55_{-0.47}^{+0.38}$ & $18.13_{-0.18}^{+0.17}$ & $17.94_{-0.25}^{+0.17}$ \\
    \hline
  \end{tabular}
  }
 \end{center}
%\vspace{1mm}
% \scriptsize{
% {\it Notes:}\\
%  $^1$For the abund.\ (in wt.\ ppm) the reported maximum values from different meteorites are given. \\
% }
\end{table}

\begin{figure}[h]
%% \vspace*{-2.0 cm}
\begin{center}
  \includegraphics[width=0.63\textwidth]{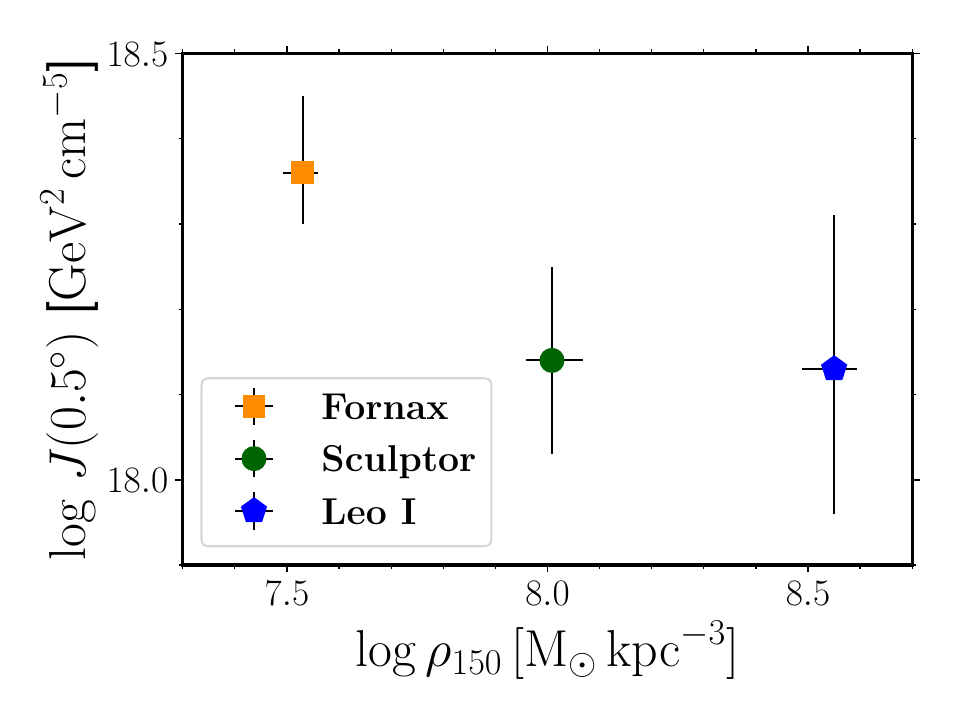}\\
   \includegraphics[width=0.63\textwidth]{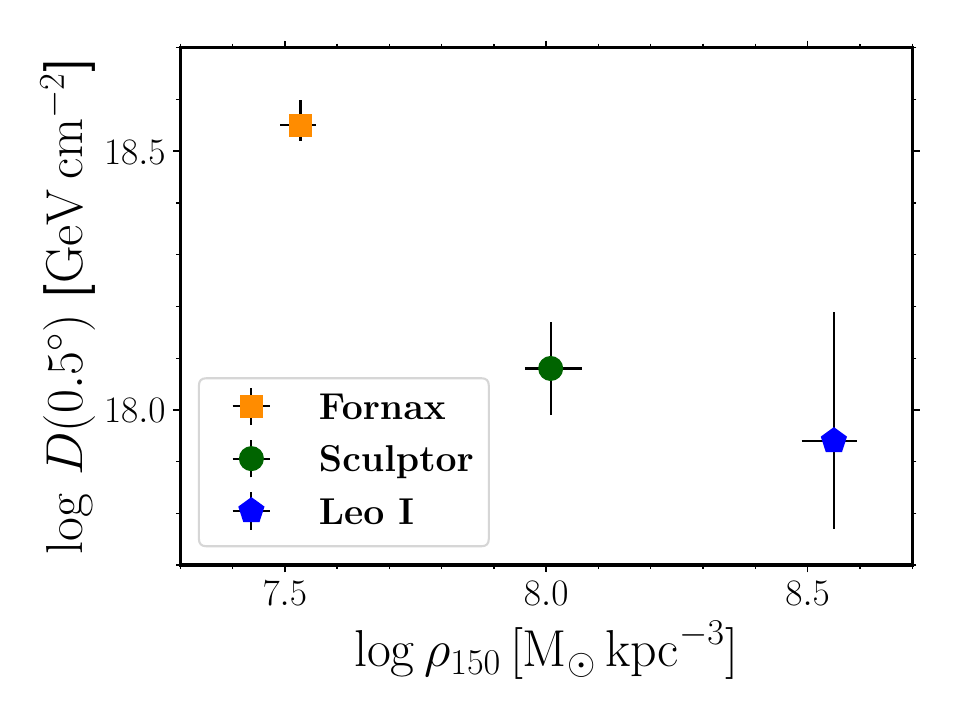} 
% \vspace*{-1.0 cm}
 \caption{$J$-factor (upper panel) and $D$-factor (lower panel),
   measured within $0.5^\circ$, as functions of central DM density
   $\rhohf$ for the Fornax, Sculptor and Leo I dSphs (see
   Table~\ref{tab:darkpar}).}
   \label{fig:jdrho150}
\end{center}
\end{figure}

The only luminous dynamical tracers in gas-poor galaxies are the
stars. If a galaxy is isolated and in equilibrium, the spatial and
kinematic distribution of its stars can be compared to stationary
dynamical models to infer the properties of the gravitational
potential and then of the total mass distribution.

Given that the Fornax, Sculptor and Leo I dSphs are satellite of the
MW, it is not guaranteed that they can be modelled as isolated and
stationary. However, there are indications that these systems are
close to equilibrium. \cite{Bat15} and \cite{DiC24} proved with
$N$-body simulations that the present-day effects of the MW tidal
field on Fornax are negligible. An analogous conclusion was reached
for Sculptor by \cite{Ior19}. Present-day tidal effects on Leo I are
expected to be negligible, because of its very large distance, though
some residual tidal disturbance in the outskirts is not excluded
because of the pericentric passage it experienced more than $1\Gyr$
ago (\cite[Bustamante-Rosell et al.\ 2021]{Bus21}, \cite[Pacucci, Ni
  \& Loeb 2023]{Pac23}).

The estimates the DM distribution in each of these dSphs were obtained
with spherically symmetric dynamical models based on distribution
functions (DFs) depending on the action integrals (\cite[Binney
  2014]{Bin14}; \cite[Vasiliev 2019]{Vas19}). The models were then
compared with observational data (stellar density distributions and
kinematics of stars) to get inference on the models' parameters.

The dynamical models and the comparison with the data are described in
detail for Fornax in \cite{Pas18}, for Sculptor in \cite{Arr_inprep},
who exploited the recent observational dataset of \cite{Tol23}, and
for Leo I in \cite{Pas24} and \cite{Pas_inprep}.  For Fornax the model
consists of a DF for the stellar component and a DF for the DM
halo. For Sculptor there are two DFs (with the same functional form as
those used in \cite[Pascale et al.\ 2024]{Pas24}) for the stars (one
for the metal-rich and the other for the metal-poor stellar
population), while the DM halo is represented by a spherical
density-potential pair, with exponentially truncated double power-law
density profile. For Leo I the model includes a DF for the stellar
component, a DF for the dark halo, and a central black hole represented
by a Keplerian potential. Remarkably, \cite{Pas24} put a $3\sigma$
upper limit $<6.76\times 10^5\Msun$ on the mass of the putative
central black hole, which would exclude the presence of a supermassive
black hole (claimed by \cite[Bustamante-Rosell et al.\ 2021]{Bus21}),
but leaving room for an intermediate mass black hole.

\section{Measurements of DM central density, $J$-factor and $D$-factor}

The results of the model-data comparison of \cite{Pas18},
\cite{Arr_inprep}, \cite{Pas24} and \cite{Pas_inprep} allowed the
authors to infer many properties of the Fornax, Sculptor and Leo I
dSphs. Here we focus on the measurements of the DM central density
$\rhohf$, $J$-factor and $D$-factor. The estimated median values of
these quantities are reported in Table~\ref{tab:darkpar}, together
with $1\sigma$ (16th-84th percentiles) confidence intervals.  In
Fig.\ \ref{fig:jdrho150} we show, for the three dSphs here considered,
the $J$-factor (upper panel) and the $D$-factor (lower panel) as
functions of $\rhohf$.  In the small sample of dSphs here considered,
the $J$- and $D$- factors tend to anticorrelate with $\rhohf$. This is
not necessarily surprising, because the $J$- and $D$- factors depend
in a complex way on the intrinsic properties of the DM density
distribution, as well as on the galaxy distance, which not only enters
through the $1/d^2$ factor, but also determines the physical aperture
size at given $\theta$ (eqs.\ \ref{eq:jfactor} and \ref{eq:dfactor}).

In the considered small sample of dSphs, Fornax is the best candidate
for indirect DM detection, having the highest $J$- and $D$-factors,
even if it has the lowest $\rhohf$ (about an order of magnitude lower
than that of Leo I). It turns out that Fornax has a combination of
distance and intrinsic properties of the DM density profile that
favours higher values of the $J$- and $D$-factors.  Among these three
galaxies, the remote dSph Leo I has the highest central DM density:
$\rhohf\approx 3.6\times 10^8\Msun\kpc^{-3}$. This is remarkable in
the context of the empirical anticorrelation between $\rhohf$ and
pericentric radius (\cite[Kaplinghat, Valli, \& Yu 2019]{Kap19}; but
see \cite[Cardona-Barrero et al.\ 2023]{Car23} for a thorough
exploration of the strength and significance of this anticorrelation),
because Leo I has one of the smallest $\rperi$ among the MW dSph
satellites.

%rho150 di Leo: 3.54828632, -0.46640789, +0.38299457) \times 10^8
%Jfactor di Leo: 18.1286896   -0.18438249  +0.16861043
%Dfactor di Leo:  17.94266045  -0.24615127  +0.17091565

%rho150 di Fornax: 3.38742789 -0.2875853 +0.24329113 x10^7
%                  0.339 -0.0288 +0.0243
%Jfactor di Fornax: 18.36 -0.09 +0.06
%Dfactor di Fornax: 18.55 -0.05 +0.03 

%SCULPTOR
%rho_150
%median: 8.01
%1 sigma (16th and 84th percentiles): -0.05, +0.06
%3 sigma (0.135th and 99.865th percentiles): -0.15, +0.16

%J(0.5 deg)
%median: 18.14
%1 sigma: -0.11, +0.11
%3 sigma: -0.31, +0.35

%D(0.5 deg)
%median: 18.08
%1 sigma: -0.09, +0.09
%3 sigma: -0.24, +0.29

%{\underline{\it Silicon carbide}}. All SiC grains in primitive meteorites are of pre-solar origin, and they are

%{\underline{\it Isotopic structures and nucleosynthesis}}. As isotopic structures are the keyy for

%\begin{discussion}

%\end{discussion}

\end{document}